\documentclass[12pt]{article}
\usepackage{graphicx}
\usepackage{color}

\def\hybrid{\topmargin 0pt      \oddsidemargin 0pt
        \headheight 0pt \headsep 0pt
       \voffset-1cm
        \textwidth 6.25in       
       \textheight 9.5in       
        \marginparwidth 0.0in
        \parskip 5pt plus 1pt   \jot = 1.5ex}
\catcode`\@=11
\def\marginnote#1{}

\newcount\hour
\newcount\minute
\newtoks\amorpm
\hour=\time\divide\hour by60
\minute=\time{\multiply\hour by60 \global\advance\minute by-\hour}
\edef\standardtime{{\ifnum\hour<12 \global\amorpm={am}%
        \else\global\amorpm={pm}\advance\hour by-12 \fi
        \ifnum\hour=0 \hour=12 \fi
        \number\hour:\ifnum\minute<10 0\fi\number\minute\the\amorpm}}
\edef\militarytime{\number\hour:\ifnum\minute<10 0\fi\number\minute}

\def\draftlabel#1{{\@bsphack\if@filesw {\let\thepage\relax
   \xdef\@gtempa{\write\@auxout{\string
      \newlabel{#1}{{\@currentlabel}{\thepage}}}}}\@gtempa
   \if@nobreak \ifvmode\nobreak\fi\fi\fi\@esphack}
        \gdef\@eqnlabel{#1}}
\def\@eqnlabel{}
\def\@vacuum{}
\def\draftmarginnote#1{\marginpar{\raggedright\scriptsize\tt#1}}

\def\draftlabel#1{{\@bsphack\if@filesw {\let\thepage\relax
   \xdef\@gtempa{\write\@auxout{\string
      \newlabel{#1}{{\@currentlabel}{\thepage}}}}}\@gtempa
   \if@nobreak \ifvmode\nobreak\fi\fi\fi\@esphack}
        \gdef\@eqnlabel{#1}}
\def\@eqnlabel{}
\def\@vacuum{}
\def\draftmarginnote#1{\marginpar{\raggedright\scriptsize\tt#1}}

\def\draft{\oddsidemargin -.5truein
        \def\@oddfoot{\sl preliminary draft \hfil
        \rm\thepage\hfil\sl\today\quad\militarytime}
        \let\@evenfoot\@oddfoot \overfullrule 3pt
        \let\label=\draftlabel
        \let\marginnote=\draftmarginnote
   \def\@eqnnum{(\theequation)\rlap{\kern\marginparsep\tt\@eqnlabel}%
\global\let\@eqnlabel\@vacuum}  }

\def\numberbysection{\@addtoreset{equation}{section}
        \def\theequation{\thesection.\arabic{equation}}}

\def\underline#1{\relax\ifmmode\@@underline#1\else
        $\@@underline{\hbox{#1}}$\relax\fi}

\def\titlepage{\@restonecolfalse\if@twocolumn\@restonecoltrue\onecolumn
     \else \newpage \fi \thispagestyle{empty}\c@page\z@
        \def\thefootnote{\fnsymbol{footnote}} }

\def\endtitlepage{\if@restonecol\twocolumn \else  \fi
        \def\thefootnote{\arabic{footnote}}
        \setcounter{footnote}{0}}  
\relax

\hybrid

\newfont{\Bbb}{msbm10 scaled 1\@ptsize00}
\newfont{\Bbbb}{msbm7 scaled 1\@ptsize00}

\newcommand{\DDD}{\raise-1pt\hbox{$\mbox{\Bbbb D}$}}

\newcommand{\UUU}{\raise-1pt\hbox{$\mbox{\Bbbb U}$}}

\newcommand{\z}{\raise-1pt\hbox{$\mbox{\Bbbb Z}$}}

\def\res{\mathop{\hbox{res}}\limits}

\def\beq{\begin{equation}}
\def\eeq{\end{equation}}
\def\p{\partial}

\begin{document}

\begin{titlepage}

\title{Time discretization of the spin Calogero-Moser model and the semi-discrete
matrix KP hierarchy}

\author{
 A.~Zabrodin\thanks{National Research University Higher School of Economics,
20 Myasnitskaya Ulitsa, Moscow 101000, Russian Federation;
Institute of Biochemical Physics of Russian Academy of Sciences,
Kosygina str. 4, 119334, Moscow, Russian Federation;
Skolkovo Institute of Science and Technology, 143026 Moscow, Russian Federation;
e-mail: zabrodin@itep.ru}
}

\date{June 2018}
\maketitle

\vspace{-7cm} \centerline{ \hfill ITEP-TH-15/18}\vspace{7cm}

\begin{abstract}

We introduce the discrete time version of the spin Calogero-Moser system.
The equations of motion follow from the dynamics of poles of rational solutions to the matrix
KP hierarchy with discrete time. The dynamics of poles is derived using the auxiliary
linear problem for the discrete flow. 

\end{abstract}

\end{titlepage}

\vspace{5mm}

\section{Introduction}

In the seminal 
paper \cite{AMM77} it was discovered that the motion of poles of rational solutions to the
Korteweg-de Vries and Boussinesq equations is given by dynamics of the 
many-body Calogero-Moser system of particles \cite{Calogero71,Calogero75,Moser75} 
with some additional restrictions in the phase space.
In \cite{Krichever78,CC77} it was shown that in the case of the Kadomtsev-Petviashvili (KP)
equation this correspondence becomes an isomorphism: poles of rational solutions to the
KP equation move exactly as Calogero-Moser particles with the interaction potential
$1/x^2$. This remarkable connection
was further generalized to elliptic solutions in \cite{Krichever80}. 
Later the correspondence between dynamics of poles of rational KP solutions and many-body
integrable systems of particles on the line was extended \cite{Shiota94} 
to the level of hierarchies. Namely, it was proved that the evolution of 
poles in $t_1=x$
with respect to the KP time $t_k$ with
$k\geq 2$ is governed by the higher Hamiltonian $H_k$ of the integrable
Calogero-Moser system. 

Similar approach to the rational solutions of the matrix KP hierarchy 
was developed in the paper \cite{PZ18}, in which the results of \cite{KBBT95}
for the matrix KP equation were extended to the whole hierarchy (see also
\cite{T11,BGK09,BZT11,BZT08,CS17}, where similar results are discussed
from other perspectives and points of view). 
The matrix extension of the KP hierarchy
is closely related to the so-called multi-component KP hierarchy 
\cite{DJKM81,KL93,Teo11,TT07}. It has been shown that
the evolution of data of the rational solutions (positions of poles $x_i$ and some 
vectors $a_{i}^{\alpha}$, $b_{i}^{\alpha}$ parameterizing rank $1$ 
matrix residues at the poles)
is isomorphic to the spin generalization of the Calogero-Moser system known as the
Gibbons-Hermsen system \cite{GH84}. 

It is natural to expect that integrable time discretization of the Calogero-Moser
system and its spin generalization can be obtained from dynamics of poles of 
rational solutions to semi-discrete soliton equations. ``Semi'' means that 
the time becomes discrete while the space variable $x$, with respect to which
one considers pole solutions, remains continuous. 
At the same time, 
it is known that integrable discretizations of soliton equations can be
regarded as belonging to the same hierarchy as their continuous counterparts.
Namely, the discrete time step is equivalent to a special shift of infinitely many
continuous hierarchical times.
This fact lies in the basis of the method of generating discrete soliton
equations developed in \cite{DJM82}. For integrable time discretization of 
many-body systems see \cite{NP94,NRK96,RS97,Suris,KWZ98}.

The equations of motion for the rational Calogero-Moser model in discrete
time $p$ have the form \cite{NP94}
\beq\label{I1}
\sum_j \frac{1}{x_i(p)-x_j(p+1)}+\sum_j \frac{1}{x_i(p)-x_j(p-1)}=
2\sum_{j\neq i} \frac{1}{x_i(p)-x_j(p)}.
\eeq
Remarkably, they coincide with the nested Bethe ansatz equations for the 
integrable quantum Gaudin magnet based on the algebra $gl(M)$, the discrete
time variable $p$ playing the role of ``level'' of the nested Bethe ansatz (in this case 
the values of $p$ are restricted to $0, 1, \ldots , M$). 

In this paper we derive equations of motion in discrete time for the
spin generalization of the Calogero-Moser model. It appears that they look like equations
(\ref{I1}) ``dressed'' by the spin variables $a_i^{\alpha}(p)$, $b_i^{\alpha}(p)$
associated with each particle with coordinate $x_i(p)$ ($\alpha =1, \ldots , N$):
\beq\label{I2}
\begin{array}{c}
\displaystyle{
\sum_j \frac{b_i^{\gamma}(p)a_j^{\gamma}(p+1)b_j^{\beta}(p+1)
a_i^{\beta}(p)}{x_i(p)-x_j(p+1)}+\sum_j \frac{b_i^{\gamma}(p)a_j^{\gamma}(p-1)
b_j^{\beta}(p-1)
a_i^{\beta}(p)}{x_i(p)-x_j(p-1)}}
\\ \\
\displaystyle{=
2\sum_{j\neq i} \frac{b_i^{\gamma}(p)a_j^{\gamma}(p)b_j^{\beta}(p)
a_i^{\beta}(p)}{x_i(p)-x_j(p)}},
\end{array}
\eeq
where the summation over repeated indices $\beta , \gamma$ is implied. 
These equations follow from the dynamics of poles of rational solutions 
to the discrete time matrix KP hierarchy. Their possible meaning from the point of view 
of Bethe ansatz is to be clarified. 

The main body of the paper starts from the multi-component KP hierarchy in the
bilinear formalism and subsequent specialization to the matrix KP hierarchy in section~2,
where the discrete time evolution is also considered. In section 3 we deal with 
pole dynamics of rational solutions to the matrix KP hierarchy. The evolution of the poles
and the vectors $a_i^{\alpha}$, $b_i^{\alpha}$ in discrete time is derived with the help
of the corresponding linear problems for the Baker-Akhiezer function and its adjoint. 
In the appendix, these linear problems are derived from the basic bilinear 
identity for the tau-function.

\section{The multi-component and matrix KP hierarchies}

We start from the multi-component KP hierarchy. Our exposition follows
\cite{Teo11,TT07}.
The multi-component KP hierarchy contains $N$ infinite sets of continuous time
variables
$$
{\bf t}=\{{\bf t}_1, {\bf t}_2, \ldots , {\bf t}_N\}, \qquad
{\bf t}_{\alpha}=\{t_{\alpha , 1}, t_{\alpha , 2}, t_{\alpha , 3}, \ldots \, \},
\qquad \alpha = 1, \ldots , N
$$
and $N$ discrete integer variables 
$
{\bf s}=\{s_1, s_2, \ldots , s_N\}
$
called charges. They are constrained by the condition
$\displaystyle{\sum_{\alpha =1}^N s_{\alpha}=0}$. In the bilinear formalism, 
the dependent variable is the tau-function $\tau ({\bf s};{\bf t})$ which depends on
the charges and the times.
The $N$-component KP hierarchy is defined as the infinite set of bilinear equations
for the tau-function that follow from the basic bilinear
identity
\beq\label{m1}
\begin{array}{l}
\displaystyle{\sum_{\gamma =1}^N \epsilon_{\alpha \gamma}({\bf s})\epsilon_{\beta \gamma}({\bf s}')
\oint_{C_{\infty}}dz \, 
z^{s_{\gamma}-s_{\gamma}'+\delta_{\alpha \gamma}+\delta_{\beta \gamma}-2}
e^{\xi ({\bf t}_{\gamma}-{\bf t}_{\gamma}', \, z)}}
\\ \\
\displaystyle{\hspace{1cm}
\cdot \tau \left ({\bf s}+{\bf e}_{\alpha}-{\bf e}_{\gamma}; {\bf t}-[z^{-1}]_{\gamma}\right )
\tau \left ({\bf s}'+{\bf e}_{\gamma}-{\bf e}_{\beta}; {\bf t}'+[z^{-1}]_{\gamma}\right )=0,
\qquad \alpha, \beta =1, \ldots , N,}
\end{array}
\eeq
valid for any ${\bf s}$, ${\bf s}'$, ${\bf t}$, ${\bf t}'$. 
Here ${\bf e}_{\alpha}$ is the vector whose $\alpha$th component is $1$ and all 
other entries are equal to zero, so for $\alpha \neq \gamma$
we have $({\bf s}+{\bf e}_{\alpha}-{\bf e}_{\gamma})_{\beta}=
s_{\beta}+\delta_{\alpha \beta}-\delta_{\gamma \beta}$. Next, we use the following standard
notation:
$$
\xi ({\bf t}_{\gamma}, z)=\sum_{k\geq 1}t_{\gamma , k}z^k,
$$
$$
\left ({\bf t}\pm [z^{-1}]_{\gamma}\right )_{\alpha k}=t_{\alpha , k}\pm
\delta_{\alpha \gamma} \frac{z^{-k}}{k}
$$
and $\epsilon_{\alpha \gamma}({\bf s})$ is the sign factor
$$
\epsilon_{\alpha \gamma}({\bf s})=\left \{
\begin{array}{ll}
\;\; (-1)^{s_{\alpha +1}+\ldots +s_{\gamma}} &\quad \mbox{if $\alpha <\gamma$}
\\
\quad 1 &\quad \mbox{if $\alpha =\gamma$}
\\
-(-1)^{s_{\gamma +1}+\ldots +s_{\alpha}} &\quad \mbox{if $\alpha >\gamma$.}
\end{array}\right. 
$$
Obviously, for any distinct $\alpha , \beta$ it holds
$\epsilon_{\alpha \beta}({\bf s})=-\epsilon_{\beta \alpha}({\bf s})$.
The integration contour $C_{\infty}$ is a big circle around $\infty$.

Along with the tau-function, an important role in the theory of integrable
hierarchies is played by the Baker-Akhiezer function. In the multi-component
KP hierarchy,
the Baker-Akhiezer function $\Psi ({\bf s}, {\bf t};z)$ and its adjoint
$\Psi ^{\dag}({\bf s}, {\bf t};z)$ are $N\! \times \! N$ matrices with components
defined by the formulae:
\beq\label{m2}
\begin{array}{l}
\displaystyle{\Psi_{\alpha \beta}({\bf s}, {\bf t};z)=
\epsilon_{\alpha \beta}({\bf s})\,
\frac{\tau \left (
{\bf s}\! +\! {\bf e}_{\alpha}\! -\! {\bf e}_{\beta}; 
{\bf t}-[z^{-1}]_{\beta}\right )}{\tau ({\bf s}; {\bf t})}\,
z^{s_{\beta}+\delta_{\alpha \beta}-1}e^{\xi ({\bf t}_{\beta}, z)},
}
\\ \\
\displaystyle{\Psi_{\alpha \beta}^{\dag}({\bf s}, {\bf t};z)=
\epsilon_{\beta \alpha}({\bf s})\,
\frac{\tau \left (
{\bf s}\! +\! {\bf e}_{\alpha}\! -\! {\bf e}_{\beta}; 
{\bf t}+[z^{-1}]_{\alpha}\right )}{\tau ({\bf s}; {\bf t})}\,
z^{-s_{\alpha}+\delta_{\alpha \beta}-1}e^{-\xi ({\bf t}_{\alpha}, z)}
}
\end{array}
\eeq
(here and below $\dag$ does not mean the Hermitian conjugation).
The complex variable $z$ is called the spectral parameter. 
Around $z=\infty$, the Baker-Akhiezer function $\Psi$ can be represented in 
the form of the series
\beq\label{m4}
\Psi_{\alpha \beta}({\bf s}, {\bf t};z)=\left (\delta_{\alpha \beta}+
\sum_{k\geq 1}\frac{w^{(k)}_{\alpha \beta}({\bf s}, 
{\bf t})}{z^k}\right )z^{s_{\beta}}e^{\xi ({\bf t}_{\beta}, z)},
\eeq
where $w^{(k)}({\bf s}, {\bf t})$ are some matrix functions.
In terms of the Baker-Akhiezer functions, the bilinear identity (\ref{m1}) can be written
as
\beq\label{m3}
\oint_{C_{\infty}}\! dz \, \Psi ({\bf s}, {\bf t};z)\Psi^{\dag} ({\bf s}', {\bf t}';z)=0.
\eeq

Another approach to the multi-component KP hierarchy is based on matrix
pseudo-differential operators. 
The hierarchy can be understood as an infinite set of 
evolution equations in the times ${\bf t}$ for matrix functions of a variable $x$. 
For example, the coefficients $w^{(k)}$ of the Baker-Akhiezer function can be taken
as such matrix functions, the evolution being $w^{(k)}(x)\to w^{(k)}(x,{\bf t})$.
In what follows we denote $\tau (x, {\bf t})$, $w^{(k)}(x,{\bf t})$ simply as 
$\tau ({\bf t})$, $w^{(k)}({\bf t})$, suppressing the dependence on $x$. 
Let us introduce the matrix pseudo-differential ``wave operator'' $W$ with matrix elements
\beq\label{m113}
W_{\alpha \beta} = \delta_{\alpha \beta}+\sum_{k\geq 1}
w^{(k)}_{\alpha \beta}({\bf t})\p_x^{-k},
\eeq
where $w^{(k)}_{\alpha \beta}({\bf t})$ are the
same matrix functions as in (\ref{m4}).
The Baker-Akhiezer function (at ${\bf s}=0$) 
can be written as a result of action of the wave operator 
to the exponential function:
\beq\label{m113a}
\Psi ({\bf t}; z)=W\exp \Bigl (xzI+\sum_{\alpha =1}^N E_{\alpha}\xi ({\bf t}_{\alpha}, z)\Bigr ),
\eeq
where $E_{\alpha}$ is the $N\! \times \! N$ matrix with the components
$(E_{\alpha})_{\beta \gamma}=\delta_{\alpha \beta}\delta_{\alpha \gamma}$. 
The adjoint Baker-Akhiezer function can be written as
\beq\label{m113b}
\Psi ^{\dag} ({\bf t}; z)=\exp \Bigl (-xzI-\sum_{\alpha =1}^N E_{\alpha}\xi ({\bf t}_{\alpha}, z)\Bigr )
W^{-1}.
\eeq
Here the operators $\p_x$ 
entering $W^{-1}$ act to the left (i.e., we define $f\p_x \equiv -\p_x f$).

It is proved in \cite{Teo11} that the Baker-Akhiezer function and its adjoint 
satisfy the linear
equations
\beq\label{m13c}
\begin{array}{l}
\,\,\,\,\, \p_{t_{\alpha , m}}\Psi ({\bf t}; z)=B_{\alpha m} \Psi ({\bf t}; z), 
\\ \\
-\p_{t_{\alpha , m}}\Psi^{\dag} ({\bf t}; z)=\Psi^{\dag} ({\bf t}; z) B_{\alpha m} , 
\end{array}
\eeq
where $B_{\alpha m}$ is the differential operator
$$
B_{\alpha m}= \Bigl (W E_{\alpha}\p_x^m W^{-1}\Bigr )_+.
$$
The notation $(\ldots )_+$ means the differential part of a pseudo-differential operator, i.e.
the sum of all terms with $\p_x^k$, where $k\geq 0$. In particular, it follows from
(\ref{m13c}) at $m=1$ that
\beq\label{m13d}
\sum_{\alpha =1}^{N}\p_{t_{\alpha , 1}}\Psi ({\bf t}; z)=\p_x \Psi ({\bf t}; z),
\eeq
so the vector field $\p_x$ can be identified with the vector field
$\sum_{\alpha }\p_{t_{\alpha , 1}}$.

The matrix KP hierarchy
results from the multicomponent KP one 
after a restriction of the time and charge
variables in the following manner:
$$
t_{\alpha , m}=t_m , \quad s_{\alpha}=0 \qquad \mbox{for each $\alpha$ and $m$}.
$$
The corresponding vector fields are related as
$\p_{t_m}=\sum_{\alpha =1}^N \p_{t_{\alpha , m}}$.
In what follows we omit the variables ${\bf s}$ in the notation for 
the tau-function and the Baker-Akhiezer functions and 
put ${\bf s}={\bf s}'=0$ in the bilinear identity, so it acquires the form
\beq\label{m5}
\sum_{\gamma =1}^N \epsilon_{\alpha \gamma}\epsilon_{\beta \gamma}
\oint_{C_{\infty}}dz \, 
z^{\delta_{\alpha \gamma}+\delta_{\beta \gamma}-2}
e^{\xi ({\bf t}_{\gamma}-{\bf t}_{\gamma}', \, z)}
\tau _{\alpha \gamma} \left ({\bf t}-[z^{-1}]_{\gamma}\right )
\tau _{\gamma \beta}\left ({\bf t}'+[z^{-1}]_{\gamma}\right )=0,
\eeq 
where 
\beq\label{m6}
\tau _{\alpha \beta}({\bf t})=\tau ({\bf e}_{\alpha}-{\bf e}_{\beta}; {\bf t})
\eeq
and
$\epsilon_{\alpha \gamma}=1$ if $\alpha \leq \gamma$, $\epsilon_{\alpha \gamma}=-1$
if $\alpha >\gamma$.
The Baker-Akhiezer function has the expansion
\beq\label{m7}
\Psi_{\alpha \beta}({\bf t};z)=\left (\delta_{\alpha \beta}+
w_{\alpha \beta}^{(1)}({\bf t})z^{-1}+O(z^{-2})\right )
e^{xz+\xi ({\bf t}, z)},
\eeq
where $\displaystyle{\xi ({\bf t}, z)=\sum_{k\geq 1}t_kz^k}$.
The coefficient $w_{\alpha \beta}^{(1)}({\bf t})$ plays an important role in what follows.
As is seen from (\ref{m2}),
\beq\label{m8}
w^{(1)}_{\alpha \beta}({\bf t})=\left \{
\begin{array}{l}
\displaystyle{\epsilon_{\alpha \beta}\, \frac{\tau_{\alpha \beta}({\bf t})}{\tau ({\bf t})}}\qquad
\,\, \mbox{if $\alpha \neq \beta$}
\\ \\
\displaystyle{-\, \frac{\p_{t_{\alpha , 1}}\tau ({\bf t})}{\tau ({\bf t})}} \qquad 
\mbox{if $\alpha = \beta$.}
\end{array}\right.
\eeq

Differentiating the bilinear
identity with respect to $t_m$ and putting ${\bf t}'={\bf t}$ after this, we 
obtain the following useful corollary:
\beq\label{m11}
\frac{1}{2\pi i}
\sum_{\gamma =1}^N \oint_{C_{\infty}}dz \, z^m \Psi_{\alpha \gamma}({\bf t}; z)
\Psi^{\dag}_{\gamma \beta}({\bf t}; z)=-\p_{t_m}w_{\alpha \beta}^{(1)}({\bf t})
\eeq
or, equivalently, 
\beq\label{m11a}
\sum_{\gamma}\res_{\infty}\Bigl (z^m \Psi_{\alpha \gamma}\Psi^{\dag}_{\gamma \beta}
\Bigr ) =-\p_{t_m}w_{\alpha \beta}^{(1)}.
\eeq

Recalling (\ref{m13d}), we can identify
$\displaystyle{
\p_x = \p_{t_1}=\sum_{\alpha =1}^N \p_{t_{\alpha , 1}}}
$.
Equations (\ref{m13c}) imply that
the Baker-Akhiezer function of the matrix KP hierarchy and its adjoint 
satisfy the linear
equations
\beq\label{m13a}
\begin{array}{l}
\,\,\,\,\, \p_{t_m}\Psi ({\bf t}; z)=B_m \Psi ({\bf t}; z), \qquad \,\,\,\, m\geq 1,
\\ \\
-\p_{t_m}\Psi^{\dag} ({\bf t}; z)=\Psi^{\dag} ({\bf t}; z) B_m , \qquad m\geq 1,
\end{array}
\eeq
where $B_m$ is the differential operator
$
B_m= \Bigl (W \p_x^m W^{-1}\Bigr )_+.
$
At $m=1$ we have $\p_{t_1}\Psi =\p_x \Psi$, so the evolution in $t_1$ is a shift of
the variable $x$:
$
w^{(k)}(x, t_1, t_2, \ldots )=w^{(k)}(x+t_1, t_2, \ldots ).
$
At $m=2$ equations (\ref{m13a}) yield the linear problems
\beq\label{m14}
\p_{t_2}\Psi = \p_x^2\Psi +V({\bf t})\Psi ,  
\eeq
\beq\label{m14a}
-\p_{t_2}\Psi^{\dag} = \p_x^2\Psi^{\dag} +\Psi^{\dag}V({\bf t})
\eeq
with
\beq\label{m15}
V({\bf t})=-2\p_x w^{(1)}({\bf t}).
\eeq

The discrete time evolution in the matrix KP hierarchy is defined as a shift 
of continuous time variables according to
the rule \cite{DJM82}
$$
\tau^p = \tau \left ({\bf t}-p\sum_{\alpha =1}^{N}[\mu^{-1}]_{\alpha}\right ), \qquad
\Psi^p=\Psi \left ({\bf t}-p\sum_{\alpha =1}^{N}[\mu^{-1}]_{\alpha};z\right ),
$$
where $p$ is the discrete time variable and $\mu$ is a continuous parameter.
Each $\mu$ corresponds to its own discrete time flow. This hierarchy is called
{\it semi-discrete} because the variable $x$ (and time $t_1$) remains continuous.
One can show, using the explicit expressions of the Baker-Akhiezer functions through the
tau-function and some corollaries of 
the bilinear identity (see the appendix) that the corresponding linear problems have the
form
\beq\label{l1}
\mu \Psi^p -\mu \Psi^{p+1}=
\p_x \Psi^p+\left (w^{(1)}(p+1)-
w^{(1)}(p)\right )\Psi^p,
\eeq
\beq\label{l2}
\mu \Psi^{\dag p} -\mu \Psi^{\dag \, p-1}=
-\p_x \Psi^{\dag p}+
\Psi^{\dag p}\left (w^{(1)}(p)-
w^{(1)}(p-1)\right ),
\eeq
or, in components,
\beq\label{l1a}
\mu \Psi_{\alpha \beta}^p -\mu \Psi_{\alpha \beta}^{p+1}=
\p_x \Psi_{\alpha \beta}^p+\sum_{\gamma}\left (w_{\alpha \gamma}^{(1)}(p+1)-
w_{\gamma \alpha}^{(1)}(p)\right )\Psi_{\gamma \beta}^p,
\eeq
\beq\label{l2a}
\mu \Psi_{\alpha \beta}^{\dag p} -\mu \Psi_{\alpha \beta}^{\dag \, p-1}=
-\p_x \Psi_{\alpha \beta}^{\dag p}+\sum_{\gamma}
\Psi_{\alpha \gamma}^{\dag p}\left (w_{\gamma \beta}^{(1)}(p)-
w_{\gamma \beta}^{(1)}(p-1)\right ).
\eeq

\section{Rational solutions to the matrix KP hierarchy and time discretization of the
Calogero-Moser model}

We are going to study solutions to the matrix KP hierarchy which are rational functions
of the variable $x$ (and, therefore, $t_1$).
For the rational solutions, the tau-function should be a polynomial in $x$:
\beq\label{r1}
\tau^p= C\prod_{i=1}^{{\cal N}} (x-x_i).
\eeq
The ${\cal N}$ roots $x_i$ (assumed to be distinct) depend 
on the times ${\bf t}$ and on the discrete variable $p$. 
They are going to be coordinates of particles in the spin Calogero-Moser system. 
Disregarding the 
dependence on ${\bf t}$, we will denote $x_i=x_i(p)$.
It is clear from (\ref{m2}) that the Baker-Akhiezer functions $\Psi$, $\Psi^\dag$
(and thus the coefficient $w^{(1)}$), 
as functions of $x$, 
have simple poles at $x=x_i$. It is shown in \cite{PZ18} that the residues 
at these poles are matrices of rank $1$. Namely, we can parametrize them
through
some column vectors ${\bf a}_i =(a_i^1, a_i^2, \ldots , a_i^N)^T$,
${\bf b}_i =(b_i^1, b_i^2, \ldots , b_i^N)^T$ ($T$ means transposition)
as follows:
\beq\label{r3}
\res_{x=x_i}\! w_{\alpha \beta}^{(1)}=-a_i^{\alpha}b_{i}^{\beta} \qquad
\mbox{or} \qquad
\res_{x=x_i}\! w^{(1)}=-{\bf a}_i {\bf b}_i^{T}.
\eeq
Note that in \cite{KBBT95} the form (\ref{r3}) was derived from some
algebro-geometric reasoning using analytic properties of the Baker-Akhiezer function
on the algebraic curve. 
For the residues of the Baker-Akhiezer functions we have \cite{PZ18}:
\beq\label{r7}
\res_{x=x_i}\! \Psi_{\alpha \beta}=e^{x_iz+\xi ({\bf t}, z)}a_i^{\alpha}c_{i}^{\beta},
\qquad
\res_{x=x_i}\! \Psi_{\alpha \beta}^{\dag}=e^{-x_iz-\xi ({\bf t}, z)}c_i^{*\alpha}b_{i}^{\beta},
\eeq
where $c_i^{\alpha}$, $c_i^{*\alpha}$ are components of some 
vectors ${\bf c}_i=(c_i^1, \ldots , c_i^N)^T$, 
${\bf c}^{*}_i=(c^{*1}_i, \ldots , c^{*N}_i)^T$.
The vectors ${\bf a}_i$, ${\bf b}_i$ depend on the times $t_k$ with $k\geq 2$ while the
vectors ${\bf c}_i$, ${\bf c}^{*}_i$ depend on the same set of times and on $z$.
The dependence of the vectors on the discrete time will be denoted as
$a_i^{\alpha}=a_i^{\alpha}(p)$, $b_i^{\alpha}=b_i^{\alpha}(p)$.
Therefore, 
we have the following representation of the Baker-Akhiezer functions:
\beq\label{r10}
\Psi _{\alpha \beta} =e^{xz+\xi ({\bf t}, z)} \left ( C_{\alpha \beta}
+\sum_{i=1}^{{\cal N}}
\frac{a_i^{\alpha}c_i^{\beta}}{x-x_i}\right ),
\eeq
\beq\label{r11}
\Psi ^{\dag} _{\alpha \beta} =e^{-xz-\xi ({\bf t}, z)} \left ( C^{-1}_{\alpha \beta}
+\sum_{i=1}^{{\cal N}}
\frac{c_i^{*\alpha}b_i^{\beta}}{x-x_i}\right ),
\eeq
where $C_{\alpha \beta}$ is a matrix of some $x$-independent coefficients. The fact that
the constant term in the adjoint Baker-Akhiezer function is the inverse matrix
$C_{\alpha \beta}^{-1}$ follows from (\ref{m113b}). 
For the matrices $w^{(1)}$ and $V=-2\p_x w^{(1)}$ we have
\beq\label{r13}
w^{(1)}_{\alpha \beta}=S_{\alpha \beta}-\sum_{i=1}^{{\cal N}} 
\frac{a_i^{\alpha}b^{\beta}_i}{x-x_i},
\qquad
V_{\alpha \beta}=-2\sum_{i=1}^{{\cal N}} \frac{a_i^{\alpha}b^{\beta}_i}{(x-x_i)^2},
\eeq
where the matrix $S$ does not depend on $x$. Tending $x\to \infty$ in (\ref{m11a}),
one concludes that $\p_{t_m}S=0$ for all $m\geq 1$, so the matrix $S$ does not depend on 
all the times.

Following \cite{PZ18,KBBT95}, we first consider the dynamics of poles 
with respect to the time $t_2$. To this end, we consider the linear problems
(\ref{m14}), (\ref{m14a}),
$$
\p_{t_2}\Psi_{\alpha \beta}=\p_{x}^2 \Psi_{\alpha \beta}-2
\sum_{i=1}^{{\cal N}}\sum_{\gamma} \frac{a_i^{\alpha}b^{\gamma}_i}{(x-x_i)^2}\, \Psi_{\gamma \beta},
$$
$$
-\p_{t_2}\Psi_{\alpha \beta}^{\dag}=\p_{x}^2 \Psi_{\alpha \beta}^{\dag}-2\sum_{\gamma}
\Psi^{\dag}_{\alpha \gamma} \sum_{i=1}^{{\cal N}}\frac{a_i^{\gamma}b^{\beta}_i}{(x-x_i)^2}
$$
and substitute here the pole ansatz for the Baker-Akhiezer functions. 
Consider first the equation for $\Psi$. 
First of all, comparing the behavior of the both sides as $x\to \infty$, we conclude that
$\p_{t_2}C_{\alpha \beta}=0$, so $C_{\alpha \beta}$ does not depend on $t_2$
(in a similar way, one can see from the higher linear problems that 
$C_{\alpha \beta}$ does not depend on all the times 
$t_k$). 
Equating coefficients at the poles at $x=x_i$ of different orders,
we get the conditions:
\begin{itemize}
\item
At $\frac{1}{(x-x_i)^3}$: \phantom{a} $b_i^{\gamma}a_i^{\gamma}=1$ or
${\bf b}_i^T {\bf a}_i =1$;
\item
At $\frac{1}{(x-x_i)^2}$: \phantom{a} $\displaystyle{
a_i^{\alpha}c_i^{\beta}\dot x_i = -2za_i^{\alpha}c_i^{\beta}-2a_i^{\alpha}\tilde b_i^{\beta}
-2\sum_{k\neq i}\frac{a_i^{\alpha}b_i^{\gamma}a_k^{\gamma}c_k^{\beta}}{x_i-x_k}}$, 
$\phantom{aaa}\tilde b_i^{\beta}=b_i^{\gamma}C_{\gamma \beta}$;
\item
At $\frac{1}{x-x_i}$: \phantom{a} $\displaystyle{\p_{t_2}(a_i^{\alpha}c_i^{\beta})=
-2\sum_{k\neq i}\frac{a_k^{\alpha}b_k^{\gamma}a_i^{\gamma}c_i^{\beta}-
a_i^{\alpha}b_i^{\gamma}a_k^{\gamma}c_k^{\beta}}{(x_i-x_k)^2}}$,
\end{itemize}
where summation over repeated Greek indices is implied and $\dot x_i=\p_{t_2}x_i$. 
The conditions coming from the
third order poles are constraints on the vectors ${\bf a}_i$, ${\bf b}_i$. 
The conditions coming from the second order poles can be written in the matrix form:
\beq\label{d1}
\sum_{k=1}^{{\cal N}}(zI-L)_{ik}c_k^{\alpha}=-\tilde b_i^{\alpha}, \qquad
L_{ik}=-\frac{\dot x_i}{2}\, \delta_{ik}-(1-\delta_{ik})\, \frac{b_i^{\gamma}
a_k^{\gamma}}{x_i-x_k},
\eeq
where $I$ is the ${\cal N}\! \times \! {\cal N}$ unity matrix. 
The matrix $L$ is going to be the Lax matrix for the 
spin Calogero-Moser system. The conditions 
at the first order poles give evolution equations in the time $t_2$.
They are written in detail in \cite{PZ18}. Similar 
calculations with the linear problem for $\Psi^{\dag}$ lead to the same
constraints ${\bf b}_i^T {\bf a}_i =1$ and to the linear equations for the vectors
${\bf c}^*_i$ with the same Lax matrix $L$:
\beq\label{d3}
\sum_{k=1}^{{\cal N}}c_k^{*\alpha}(zI-L)_{ki}=\tilde a_i^{\alpha}, \qquad
\tilde a_i^{\alpha}=C^{-1}_{\alpha \gamma}a_i^{\gamma}.
\eeq
For completeness, we give here the equations of motion 
in the time $t_2$ (see \cite{PZ18} for details):
\beq\label{eqm1}
\dot a_i^{\alpha}=-2\sum_{k\neq i}\frac{b_k^{\gamma}
a_i^{\gamma}a_k^{\alpha}}{(x_i-x_k)^2}, \qquad
\dot b_i^{\alpha}=2\sum_{k\neq i}\frac{b_i^{\gamma}
a_k^{\gamma}b_k^{\alpha}}{(x_i-x_k)^2},
\eeq
\beq\label{eqm2}
\ddot x_i=-8\sum_{k\neq i}\frac{b_i^{\gamma}
a_k^{\gamma}b_k^{\gamma '}a_i^{\gamma '}}{(x_i-x_k)^3}.
\eeq

Now we turn to the discrete time evolution. The Baker-Akhiezer functions are
\beq\label{dis1}
\Psi_{\alpha \beta}^p=\Bigl (1-\frac{z}{\mu}\Bigr )^p e^{xz+\xi ({\bf t}, z)}
\left ( C_{\alpha \beta}+\sum_i \frac{a_i^{\alpha}(p)c_i^{\beta}(p)}{x-x_i(p)}\right ),
\eeq
\beq\label{dis2}
\Psi_{\alpha \beta}^{\dag p}=\Bigl (1-\frac{z}{\mu}\Bigr )^{-p} e^{-xz-\xi ({\bf t}, z)}
\left ( C^{-1}_{\alpha \beta}+
\sum_i \frac{c_i^{*\alpha}(p)b_i^{\beta}(p)}{x-x_i(p)}\right ).
\eeq
We should substitute them into the linear problems (\ref{l1a}), (\ref{l2a}) and 
compare the coefficients at the poles at $x=x_i(p)$ and $x=x_i(p+1)$. Note that the
constant term $S_{\alpha \beta}$ in $w^{(1)}_{\alpha \beta}(p)$ cancels in the combination
$w^{(1)}_{\alpha \beta}(p+1)-w^{(1)}_{\alpha \beta}(p)$ because 
the shift $p\to p+1$ is equivalent to a shift of times and
$S_{\alpha \beta}$ 
does not depend on them. The cancellation of poles gives the following conditions:
\begin{itemize}
\item
At $\frac{1}{(x-x_i(p))^2}$: \phantom{a} $b_i^{\gamma}(p)a_i^{\gamma}(p)=1$;
\item
At $\frac{1}{x-x_i(p+1)}$:  $$
(z-\mu )a_i^{\alpha}(p+1)c_i^{\beta}(p+1)=-a_i^{\alpha}(p+1)\tilde b_i^{\beta}(p+1)-
\sum_j\frac{a_i^{\alpha}(p+1)b_i^{\gamma}(p+1)a_j^{\gamma}(p)c_j^{\beta}(p)}{x_i(p+1)
-x_j(p)};$$
\item
At $\frac{1}{x-x_i(p)}$: 
$$
(z-\mu )a_i^{\alpha}(p)c_i^{\beta}(p)+a_i^{\alpha}(p)\tilde b_i^{\beta}(p)
-\sum_j\frac{a_j^{\alpha}(p+1)b_j^{\gamma}(p+1)a_i^{\gamma}(p)c_i^{\beta}(p)}{x_i(p)
-x_j(p+1)}
$$
$$
+\sum_{j\neq i}\frac{a_i^{\alpha}(p)b_i^{\gamma}(p)
a_j^{\gamma}(p)c_j^{\beta}(p)}{x_i(p) -x_j(p)}
+\sum_{j\neq i}\frac{a_j^{\alpha}(p)b_j^{\gamma}(p)
a_i^{\gamma}(p)c_i^{\beta}(p)}{x_i(p) -x_j(p)}=0.
$$
\end{itemize}
Similar conditions follow from the linear problem for $\Psi^{\dag}$:
\begin{itemize}
\item
At $\frac{1}{(x-x_i(p))^2}$: \phantom{a} $b_i^{\gamma}(p)a_i^{\gamma}(p)=1$;
\item
At $\frac{1}{x-x_i(p-1)}$:  $$
(z-\mu )c_i^{*\alpha}(p-1)b_i^{\beta}(p-1)=\tilde a_i^{\alpha}(p-1) b_i^{\beta}(p-1)+
\sum_j\frac{c_j^{*\alpha}(p)b_j^{\gamma}(p)a_i^{\gamma}(p-1)b_i^{\beta}(p-1)}{x_i(p-1)
-x_j(p)};$$
\item
At $\frac{1}{x-x_i(p)}$: 
$$
(z-\mu )c_i^{*\alpha}(p)b_i^{\beta}(p)-\tilde a_i^{\alpha}(p) b_i^{\beta}(p)
+\sum_j\frac{c_i^{*\alpha}(p)b_i^{\gamma}(p)a_j^{\gamma}(p-1)b_j^{\beta}(p-1)}{x_i(p)
-x_j(p-1)}
$$
$$
-\sum_{j\neq i}\frac{c_i^{*\alpha}(p)b_i^{\gamma}(p)a_j^{\gamma}(p)
b_j^{\beta}(p)}{x_i(p) -x_j(p)}
+\sum_{j\neq i}\frac{c_j^{*\alpha}(p)b_j^{\gamma}(p)a_i^{\gamma}(p)
b_i^{\beta}(p)}{x_j(p) -x_i(p)}=0.
$$
\end{itemize}
The condition at the second order pole is the same constraint as before. 
Introduce the matrices
\beq\label{dis3}
L_{ij}(p)=-\delta_{ij}\frac{\dot x_i(p)}{2}-(1-\delta_{ij})\,
\frac{b_i^{\gamma}(p)a_j^{\gamma}(p)}{x_i(p)-x_j(p)}
\eeq
(the same Lax matrix as in (\ref{d1})) and
\beq\label{dis4}
M_{ij}(p)=\frac{b_i^{\gamma}(p+1)a_j^{\gamma}(p)}{x_i(p+1)-x_j(p)},
\eeq
then the above conditions coming from the first order poles at
$x_i(p)$ and $x_i(p \pm 1)$ can be written as
\beq\label{dis5}
\left \{
\begin{array}{l}
\displaystyle{(z-\mu )c_i^{\beta}(p+1)=-\tilde b_i^{\beta}(p+1)-
\sum_j M_{ij}(p)c_j^{\beta}(p)}
\\ \\
\displaystyle{ a_i^{\alpha}(p)\underbrace{\left [\sum_j\Bigl (z\delta_{ij}-L_{ij}(p)\Bigr )
c_j^{\beta}(p)+\tilde b_i^{\beta}(p)\right ]}_{=0}}
\\ \\
\displaystyle{
\phantom{aaaaaaaaaaaa}
+c_i^{\beta}(p)\left [ \sum_j a_j^{\alpha}(p+1)M_{ji}(p) +\sum_j a_j^{\alpha}(p)L_{ji}(p)
-\mu a_i^{\alpha}(p)\right ]=0,}
\end{array}
\right.
\eeq
\beq\label{dis6}
\left \{
\begin{array}{l}
\displaystyle{(z-\mu )c_i^{*\alpha}(p-1)=\tilde a_i^{\alpha}(p-1)-
\sum_j c_j^{*\alpha}(p)M_{ji}(p-1)}
\\ \\
\displaystyle{ b_i^{\beta}(p)\underbrace{\left [\sum_j c_j^{*\alpha}(p)
\Bigl (z\delta_{ij}-L_{ji}(p)\Bigr )
-\tilde a_i^{\alpha}(p)\right ]}_{=0}}
\\ \\
\displaystyle{
\phantom{aaaaaaaaaaaa}
+c_i^{*\alpha}(p)\left [ \sum_j M_{ij}(p-1)b_j^{\beta}(p-1) +\sum_j L_{ij}(p)b_j^{\beta}(p)
-\mu b_i^{\beta}(p)\right ]=0.}
\end{array}
\right.
\eeq
The first brackets in the second equations vanish because of (\ref{d1}), (\ref{d3}).
Introduce the ${\cal N}$-component column vectors
${\bf C}^{\alpha}=(c_1^{\alpha}, \ldots , c_{{\cal N}}^{\alpha})^T$,
${\bf C}^{*\alpha}=(c_1^{*\alpha}, \ldots , c_{{\cal N}}^{*\alpha})^T$,
${\bf A}^{\alpha}=(a_1^{\alpha}, \ldots , a_{{\cal N}}^{\alpha})^T$,
${\bf B}^{\alpha}=(b_1^{\alpha}, \ldots , b_{{\cal N}}^{\alpha})^T$ and
$\tilde {\bf A}^{\alpha}=(\tilde a_1^{\alpha}, \ldots , \tilde a_{{\cal N}}^{\alpha})^T$,
$\tilde {\bf B}^{\alpha}=(\tilde b_1^{\alpha}, \ldots , \tilde b_{{\cal N}}^{\alpha})^T$.
In this notation, the above equations are rewritten as linear equations for the
vectors ${\bf A}^{\alpha}$ and ${\bf B}^{\beta}$,
\beq\label{dis7}
\left \{
\begin{array}{l}
{\bf A}^{\alpha T}(p+1)M(p)+{\bf A}^{\alpha T}(p)L(p)=\mu {\bf A}^{\alpha T}(p)
\\ \\
M(p-1){\bf B}^{\beta}(p-1)+L(p){\bf B}^{\beta}(p)=\mu {\bf B}^{\beta}(p),
\end{array}
\right.
\eeq
and linear equations for the vectors ${\bf C}^{\alpha}$ and ${\bf C}^{*\alpha}$:
\beq\label{dis8}
\left \{
\begin{array}{l}
(z-\mu ){\bf C}^{\beta}(p+1)=-\tilde {\bf B}^{\beta}(p+1)-M(p){\bf C}^{\beta}(p)
\\ \\
(z-\mu ){\bf C}^{\beta}(p)=-\tilde {\bf B}^{\beta}(p)+L(p){\bf C}^{\beta}(p)-
\mu {\bf C}^{\beta}(p),
\end{array}
\right.
\eeq
\beq\label{dis9}
\left \{
\begin{array}{l}
(z-\mu ){\bf C}^{*\alpha T}(p-1)=\tilde {\bf A}^{\alpha T}(p-1)-{\bf C}^{*\alpha T}(p)M(p-1)
\\ \\
(z-\mu ){\bf C}^{*\alpha T}(p)=\tilde {\bf A}^{\alpha T}(p)+{\bf C}^{*\alpha T}(p)L(p)-
\mu {\bf C}^{*\alpha T}(p).
\end{array}
\right.
\eeq
From equations (\ref{dis8}) we have
\beq\label{dis10}
M(p){\bf C}^{\beta}(p)+\Bigl ( L(p+1)-\mu I\Bigr ){\bf C}^{\beta}(p+1)=0
\eeq
while from (\ref{dis9}) we have
\beq\label{dis11}
{\bf C}^{*\alpha T}(p+1)M(p)+{\bf C}^{*\alpha T}(p)\Bigl (L(p)-\mu I\Bigr )=0.
\eeq
Substituting equations (\ref{dis8}) into (\ref{dis10}) and equations 
(\ref{dis9}) into (\ref{dis11}), we obtain
$$
M(p)\Bigl (-\tilde {\bf B}^{\beta}(p)+L(p){\bf C}^{\beta}(p)-\mu {\bf C}^{\beta}(p)\Bigr )
-\Bigl ( L(p+1)\! -\! 
\mu I\Bigr )\Bigl (\tilde {\bf B}^{\beta}(p+1)+M(p){\bf C}^{\beta}(p)\Bigr )=0,
$$ 
$$
\begin{array}{l}
\Bigl (\tilde {\bf A}^{\alpha T}(p)+{\bf C}^{*\alpha T}(p)L(p)-\mu {\bf C}^{*\alpha T}(p)
\Bigr ) M(p-1)
\\ \\
\phantom{aaaaaaaaaaaaaa}
+\Bigl (\tilde {\bf A}^{\alpha T}(p-1)-{\bf C}^{*\alpha T}(p)M(p-1)\Bigr )
\Bigl ( L(p-1)\! -\! \mu I\Bigr )=0,
\end{array}
$$
or, after simplifying and taking into account equations (\ref{dis7}),
$$
\Bigl (M(p)L(p)-L(p+1)M(p)\Bigr ) {\bf C}^{\beta}(p)=0,
$$
$$
{\bf C}^{*\alpha T}(p+1)\Bigl (M(p)L(p)-L(p+1)M(p)\Bigr )=0.
$$
This implies the consistency condition 
\beq\label{dis12}
L(p+1)M(p)=M(p)L(p)
\eeq
or $L(p+1)=M(p)L(p)M^{-1}(p)$ which is the discrete Lax equation. 
One can show by a direct calculation that it holds provided the equations
(\ref{dis7}) are satisfied, which are equations of motion for the poles $x_i$ and
components of the vectors ${\bf a}_i$, ${\bf b}_i$. Let us write down these
equations in detail:
\beq\label{dis13}
\sum_j \frac{b_i^{\gamma}(p)a_{j}^{\gamma}(p-1)b_j^{\beta}(p-1)}{x_i(p)-
x_j(p-1)}=\frac{\dot x_i(p)}{2}\, b_i^{\beta}(p)+
\sum_{j\neq i}\frac{b_i^{\gamma}(p)a_{j}^{\gamma}(p)b_j^{\beta}(p)}{x_i(p)-
x_j(p)}+\mu  b_i^{\beta}(p),
\eeq
\beq\label{dis14}
\sum_j \frac{a_j^{\alpha}(p+1)b_{j}^{\gamma}(p+1)a_i^{\gamma}(p)}{x_j(p+1)-
x_i(p)}=\frac{\dot x_i(p)}{2}\, a_i^{\alpha}(p)+
\sum_{j\neq i}\frac{a_j^{\alpha}(p)b_{j}^{\gamma}(p)a_i^{\gamma}(p)}{x_j(p)-
x_i(p)}+\mu  a_i^{\alpha}(p).
\eeq
Multiply the first equation by $a_i^{\beta}(p)$ and sum over $\beta$, then 
multiply the second equation by $b_i^{\alpha}(p)$, sum over $\alpha$ and take into 
account the constraint $b_i^{\gamma}a_i^{\gamma}=1$.
Subtracting the resulting equations, we eliminate $\dot x_i (p)$ and obtain the 
equations of motion (\ref{I2}):
\beq\label{dis15}
\begin{array}{c}
\displaystyle{
\sum_j \frac{b_i^{\gamma}(p)a_j^{\gamma}(p+1)b_j^{\beta}(p+1)
a_i^{\beta}(p)}{x_i(p)-x_j(p+1)}+\sum_j \frac{b_i^{\gamma}(p)a_j^{\gamma}(p-1)
b_j^{\beta}(p-1)
a_i^{\beta}(p)}{x_i(p)-x_j(p-1)}}
\\ \\
\displaystyle{=
2\sum_{j\neq i} \frac{b_i^{\gamma}(p)a_j^{\gamma}(p)b_j^{\beta}(p)
a_i^{\beta}(p)}{x_i(p)-x_j(p)}}.
\end{array}
\eeq
Summing the resulting equations, we obtain an expression for $\dot x_i(p)$:
\beq\label{dis16}
\dot x_i(p)=\sum_j \frac{b_i^{\gamma}(p)a_j^{\gamma}(p-1)
b_j^{\beta}(p-1)
a_i^{\beta}(p)}{x_i(p)-x_j(p-1)}-
\sum_j \frac{b_i^{\gamma}(p)a_j^{\gamma}(p+1)b_j^{\beta}(p+1)
a_i^{\beta}(p)}{x_i(p)-x_j(p+1)}-2\mu .
\eeq

Equations (\ref{dis13}), (\ref{dis14}) are not closed since they contain 
$\dot x_i$, the derivative of $x_i$ with respect to the ``wrong'' time $t_2$.
However, it is possible to make them closed by differentiating both sides with respect to 
$t_2$ and using the equations (\ref{eqm1}), (\ref{eqm2}) for $\dot a_i^{\alpha}$,
$\dot b_i^{\alpha}$, $\ddot x_i$ and equations (\ref{dis13}), (\ref{dis14}) themselves
for $\dot a_i^{\alpha}\dot x_i$, $\dot b_i^{\alpha}\dot x_i$. After some mysterious 
cancellations one arrives in this way at the following relatively compact equations:
{\small
$$
\sum_{j,k}\frac{b_i^{\gamma}(p)a_j^{\gamma}(p\! -\! 1)b_j^{\gamma '}(p\! -\! 1)
a_k^{\gamma '}(p\! -\! 2)b_k^{\beta}(p\! -\! 2)}{(x_j(p-1)\! -\! x_i(p))^2
(x_k(p-2)\! -\! x_j(p-1))}
+\sum_{j,k}\frac{b_i^{\gamma}(p)a_k^{\gamma}(p)b_k^{\gamma '}(p)
a_j^{\gamma '}(p\! -\! 1)b_j^{\beta}(p\! -\! 1)}{(x_j(p-1)\! -\! x_i(p))^2
(x_k(p)\! -\! x_j(p-1))}
$$
\beq\label{dis18}
+\sum_{j\neq k}\frac{b_i^{\gamma}(p)b_k^{\gamma '}(p\! -\! 1)a_j^{\gamma '}(p\! -\! 1)
a_k^{\gamma }(p\! -\! 1)b_j^{\beta}(p\! -\! 1)+
b_i^{\gamma}(p)b_j^{\gamma '}(p\! -\! 1)a_k^{\gamma '}(p\! -\! 1)
a_j^{\gamma }(p\! -\! 1)b_k^{\beta}(p\! -\! 1)}{(x_j(p-1)\! -\! x_i(p))^2
(x_i(p)\! -\! x_k(p-1))}=0,
\eeq
}
{\small
$$
\sum_{j,k}\frac{b_j^{\gamma}(p\! +\! 1)a_i^{\gamma}(p)b_k^{\gamma '}(p\! +\! 2)
a_j^{\gamma '}(p\! +\! 1)a_k^{\alpha}(p\! +\! 2)}{(x_j(p+1)\! -\! x_i(p))^2
(x_k(p+2)\! -\! x_j(p+1))}
+\sum_{j,k}\frac{b_j^{\gamma}(p\! +\! 1)a_k^{\gamma}(p)b_k^{\gamma '}(p)
a_i^{\gamma '}(p)a_j^{\alpha}(p\! +\! 1)}{(x_j(p+1)\! -\! x_i(p))^2
(x_k(p)\! -\! x_j(p+1))}
$$
\beq\label{dis19}
+\sum_{j\neq k}\frac{b_k^{\gamma}(p\! +\! 1)a_j^{\gamma}(p\! +\! 1)b_j^{\gamma '}(p\! +\! 1)
a_i^{\gamma '}(p)a_k^{\alpha}(p\! +\! 1)+
b_j^{\gamma}(p\! +\! 1)a_k^{\gamma}(p\! +\! 1)b_k^{\gamma '}(p\! +\! 1)
a_i^{\gamma '}(p)a_j^{\alpha}(p\! +\! 1)}{(x_j(p+1)\! -\! x_i(p))^2
(x_i(p)\! -\! x_k(p+1))}=0.
\eeq
}

At last, let us discuss the continuum limit of equations (\ref{dis15}),
(\ref{dis13}), (\ref{dis14}). We expect that the continuous time flow $t$
corresponding to $p$ tends to $t_2$.
We set $x_i(p)=\lambda p +y_i(p)$ and expand
$x_i(p\pm 1)=\pm \lambda +x_i(p)\pm \varepsilon \p _t y_i +\frac{\varepsilon ^2}{2}\, 
\p ^2_t y_i
+O(\varepsilon ^3)$, 
$a_i^{\alpha}(p\pm 1)=a_i^{\alpha}\pm \varepsilon \p _t a_i^{\alpha}+
O(\varepsilon ^2)$, $b_i^{\alpha}(p\pm 1)=b_i^{\alpha}\pm \varepsilon \p _t b_i^{\alpha}+
O(\varepsilon ^2)$, 
where 
$\lambda =O(\sqrt{\varepsilon})$. Separating the terms with $j=i$ in the first line
of (\ref{dis15}), we expand this equation up to the first non-vanishing order
$O(\varepsilon )$ as $\varepsilon \to 0$ and obtain
\beq\label{dis17}
\p _t^2 y_i=-2g \sum_{j\neq i}
\frac{b_i^{\gamma}a_j^{\gamma}
b_j^{\beta}a_i^{\beta}}{(y_i-y_j)^3}, \qquad g=\lambda^4 /\varepsilon ^2 =O(1),
\eeq
which are equations of motion for the continuous time spin Calogero-Moser system.
We see that the coupling constant depends on the way of tending the time step to zero.
Comparison with (\ref{eqm2}) shows that one should put $g=4$, i.e., 
$\lambda^4 =4\varepsilon ^2$, then in the limit one has $\p _t^2 y_i =\ddot y_i$.

The continuum limit of equations (\ref{dis13}), (\ref{dis14}) is a little bit 
more tricky. Expanding (\ref{dis13}) as $\lambda , \varepsilon \to 0$, we obtain:
$$
\frac{b_i^{\beta}}{\lambda}-\frac{\varepsilon}{\lambda^2}\, b_i^{\beta}\p _t y_i
-\frac{\varepsilon}{\lambda}\, b_i^{\gamma}\p _t a_i^{\gamma}b_i^{\beta}
-\frac{\varepsilon}{\lambda}\, \p _t b_i^{\beta}-\lambda \sum_{j\neq i}
\frac{b_i^{\gamma}a_j^{\gamma}b_j^{\beta}}{(y_i-y_j)^2} +O(\varepsilon )=
\frac{\dot y_i}{2}\, b_i^{\beta} +\mu b_i^{\beta}.
$$
Comparison of the leading terms gives $\lambda =\mu^{-1}$. The next order terms
give
$$
\p _t y_i =-\frac{\lambda^2}{2\varepsilon}\, \dot y_i -\lambda b_i^{\gamma}
\p_t a_i^{\gamma}+O(\varepsilon ).
$$
In order to make the $t$-flow identical to the $t_2$-flow in the limit, one should
require $\lambda^2 =-2\varepsilon$ which agrees with the condition $\lambda^4 =4\varepsilon ^2$
mentioned above. Then in the order $O(\lambda )$ we obtain
$$
\p_t b_i^{\beta}=2\sum_{j\neq i}\frac{b_i^{\gamma}a_j^{\gamma}b_j^{\beta}}{(y_i-y_j)^2},
$$
which is the second equation in (\ref{eqm1}). The first one is obtained in a similar
way from equation (\ref{dis14}). 

\section{Conclusion}

To conclude, we have derived the discrete time equations of motion for the characteristic 
data of rational solutions
to the semi-discrete matrix KP hierarchy -- positions of ${\cal N}$ poles $x_i$ 
and vectors $a_i^{\alpha}$, $b_i^{\alpha}$ parametrizing rank $1$ matrix residues
at the poles. These equations define integrable time discretization of the spin
Calogero-Moser ${\cal N}$-particle system (which is known as the Gibbons-Hermsen system).
They look like equations of motion for the discrete time spinless Calogero-Moser system
``dressed'' by scalar products of the vectors $a_i^{\alpha}$, $b_j^{\alpha}$.
The continuum limit of these equation coincides with the equations of motion for the 
Gibbons-Hermsen system. The main technical tool for the derivation of the discrete time
equations of motion is the linear problems for the Baker-Akhiezer functions of the 
semi-discrete matrix KP hierarchy, which are obtained as corollaries of the basic
bilinear identity for the tau-function.

It would be interesting to extend the results of this work to elliptic 
solutions to the semi-discrete matrix KP hierarchy. We expect that the method of
the paper \cite{Krichever80} is applicable in this case and the resulting equations
of motion have a similar structure, with the simple pole function $(x-x_i)^{-1}$ being
replaced by the Weierstrass $\zeta$-function $\zeta (x-x_i)$.

\section*{Appendix}
\def\theequation{A\arabic{equation}}
\setcounter{equation}{0}

\subsection*{Some corollaries of the bilinear identity}

Here we list some corollaries of the basic bilinear identity (\ref{m5}) which are
used below in the appendix for the derivation of the linear 
problems (\ref{l1a}), (\ref{l2a}).

Differentiating the bilinear identity (\ref{m5})  
with respect to $t_{\gamma , 1}$ and setting
${\bf t}'={\bf t}-[\mu^{-1}]_{\beta}$, we obtain, after calculating the residues,
for any distinct $\alpha , \beta , \gamma$:
\beq\label{A1}
\tau_{\alpha \beta}({\bf t}\! -\! [\mu^{-1}]_{\beta})\p_{t_{\gamma , 1}}\tau ({\bf t})-
\tau ({\bf t})\p_{t_{\gamma , 1}}\tau_{\alpha \beta}({\bf t}\! -\! [\mu^{-1}]_{\beta})
+\frac{\epsilon_{\alpha \gamma}
\epsilon_{\gamma \beta}}{\epsilon_{\alpha \beta}}\,
\tau_{\alpha \gamma}({\bf t})\tau_{\gamma \beta}({\bf t}\! -\! [\mu^{-1}]_{\beta})=0
\eeq
(no summation over repeated indices!).
Differentiating (\ref{m5}) with respect to $t_{\beta , 1}$ and setting
${\bf t}'={\bf t}-[\mu^{-1}]_{\alpha}-[\nu^{-1}]_{\beta}$,
we obtain, for any distinct $\alpha , \beta$:
\beq\label{A2}
\begin{array}{c}
\p_{t_{\beta , 1}}\tau_{\alpha \beta}({\bf t}-[\nu^{-1}]_{\beta})
\tau ({\bf t}-[\mu^{-1}]_{\alpha})-
\p_{t_{\beta , 1}}\tau ({\bf t}-[\mu^{-1}]_{\alpha})
\tau_{\alpha \beta}({\bf t}-[\nu^{-1}]_{\beta})
\\ \\
+\nu \tau_{\alpha \beta}({\bf t}-[\nu^{-1}]_{\beta})
\tau ({\bf t}-[\mu^{-1}]_{\alpha})
-\nu \tau_{\alpha \beta}({\bf t})\tau ({\bf t}-[\mu^{-1}]_{\alpha}-[\nu^{-1}]_{\beta})=0.
\end{array}
\eeq
In a similar way, differentiating (\ref{m5}) with respect to $t_{\alpha , 1}$ and setting
${\bf t}'={\bf t}-[\mu^{-1}]_{\alpha}-[\nu^{-1}]_{\beta}$,
we obtain, for any distinct $\alpha , \beta$:
\beq\label{A3}
\begin{array}{c}
\p_{t_{\alpha , 1}}\tau_{\alpha \beta}({\bf t}-[\nu^{-1}]_{\beta})
\tau ({\bf t}-[\mu^{-1}]_{\alpha})-
\p_{t_{\alpha , 1}}\tau ({\bf t}-[\mu^{-1}]_{\alpha})
\tau_{\alpha \beta}({\bf t}-[\nu^{-1}]_{\beta})
\\ \\
-\mu \tau_{\alpha \beta}({\bf t}-[\nu^{-1}]_{\beta})
\tau ({\bf t}-[\mu^{-1}]_{\alpha})
+\mu \tau ({\bf t})\tau_{\alpha \beta}({\bf t}-[\mu^{-1}]_{\alpha}-[\nu^{-1}]_{\beta})=0.
\end{array}
\eeq
Differentiating (\ref{m5}) at $\beta =\alpha$ 
with respect to $t_{\gamma , 1}$ ($\gamma \neq \alpha$) and setting
${\bf t}'={\bf t}-[\mu^{-1}]_{\alpha}$,
we obtain, for any distinct $\alpha , \gamma$:
\beq\label{A4}
\p_{t_{\gamma , 1}}\tau ({\bf t}-[\mu^{-1}]_{\alpha})\tau ({\bf t})-
\p_{t_{\gamma , 1}}\tau ({\bf t})\tau ({\bf t}-[\mu^{-1}]_{\alpha})+
\mu^{-1}\tau_{\alpha \gamma}({\bf t})\tau_{\gamma \alpha}({\bf t}-[\mu^{-1}]_{\alpha})=0.
\eeq

Setting ${\bf t}'={\bf t}+[\mu ^{-1}]-[\nu ^{-1}]_{\beta}$, where we abbreviate
$\displaystyle{[\mu ^{-1}]\equiv \sum_{\gamma =1}^{N}[\mu ^{-1}]_{\gamma}}$, 
we represent the bilinear identity (\ref{m5}) at $\alpha \neq \beta$ in the form
$$
\epsilon_{\beta \alpha}\oint_{C_{\infty}}\! dz  z^{-1}\Bigl (1-\frac{z}{\mu}\Bigr )
\tau \Bigl ({\bf t}-[z^{-1}]_{\alpha}\Bigr )
\tau_{\alpha \beta} \Bigl ({\bf t}+ 
[\mu ^{-1}]-[\nu ^{-1}]_{\beta} +[z^{-1}]_{\alpha}\Bigr )
$$
$$
+\, \epsilon_{\alpha \beta}\oint_{C_{\infty}}\! dz z^{-1}
\frac{1-\frac{z}{\mu}}{1-\frac{z}{\nu}}\,
\tau_{\alpha \beta} \Bigl ({\bf t}-[z^{-1}]_{\beta}\Bigr )
\tau \Bigl ({\bf t}+ 
[\mu ^{-1}]-[\nu ^{-1}]_{\beta} +[z^{-1}]_{\beta}\Bigr )
$$
$$
+\, \sum_{\gamma \neq \alpha , \beta}
\epsilon_{\alpha \gamma}\epsilon_{\beta \gamma}\oint_{C_{\infty}}\! dz z^{-2}
\Bigl (1-\frac{z}{\mu}\Bigr ) 
\tau_{\alpha \gamma} \Bigl ({\bf t}-[z^{-1}]_{\gamma}\Bigr )
\tau_{\gamma \beta} \Bigl ({\bf t}+ 
[\mu ^{-1}]-[\nu ^{-1}]_{\beta} +[z^{-1}]_{\gamma}\Bigr )=0.
$$
Calculating the residues and multiplying by $\mu$, we obtain:
\beq\label{A5}
\begin{array}{c}
\epsilon_{\beta \alpha}\mu \tau ({\bf t}) \tau_{\alpha \beta} \Bigl ({\bf t}+ 
[\mu ^{-1}]-[\nu ^{-1}]_{\beta}\Bigr )+
\epsilon_{\beta \alpha}\p_{t_{\alpha , 1}}\tau ({\bf t})
\tau_{\alpha \beta} \Bigl ({\bf t}+ [\mu ^{-1}]-[\nu ^{-1}]_{\beta}\Bigr )
\\ \\
-\epsilon_{\beta \alpha}\tau ({\bf t})\p_{t_{\alpha , 1}}
\tau_{\alpha \beta} \Bigl ({\bf t}+ [\mu ^{-1}]-[\nu ^{-1}]_{\beta}\Bigr )
\\ \\
+\epsilon_{\alpha \beta}(\mu -\nu )
\tau_{\alpha \beta} \Bigl ({\bf t}-[\nu ^{-1}]_{\beta}\Bigr )
\tau \Bigl ({\bf t}+[\mu ^{-1}]\Bigr )+
\epsilon_{\alpha \beta}\nu \tau_{\alpha \beta} ({\bf t})
\tau \Bigl ({\bf t}+ [\mu ^{-1}]-[\nu ^{-1}]_{\beta}\Bigr )
\\ \\
\displaystyle{-\sum_{\gamma \neq \alpha , \beta}
\epsilon_{\alpha \gamma}\epsilon_{\beta \gamma}\tau_{\alpha \gamma} ({\bf t})
\tau_{\gamma \beta} \Bigl ({\bf t}+ [\mu ^{-1}]-[\nu ^{-1}]_{\beta}\Bigr )=0.
}
\end{array}
\eeq

Setting ${\bf t}'={\bf t}+[\mu ^{-1}]-[\nu ^{-1}]_{\alpha}$, 
we represent the bilinear identity (\ref{m5}) at $\beta =\alpha$ in the form
$$
\oint_{C_{\infty}}\! dz 
\frac{1-\frac{z}{\mu}}{1-\frac{z}{\nu}}\,
\tau \Bigl ({\bf t}-[z^{-1}]_{\alpha}\Bigr )
\tau \Bigl ({\bf t}+ 
[\mu ^{-1}]-[\nu ^{-1}]_{\alpha} +[z^{-1}]_{\alpha}\Bigr )
$$
$$
+\sum_{\gamma \neq \alpha}\oint_{C_{\infty}}\! dz z^{-2}
\Bigl ( 1-\frac{z}{\mu}\Bigr )\, \tau_{\alpha \gamma} 
\Bigl ({\bf t}-[z^{-1}]_{\gamma}\Bigr )
\tau_{\gamma \alpha} \Bigl ({\bf t}+ 
[\mu ^{-1}]-[\nu ^{-1}]_{\alpha} +[z^{-1}]_{\gamma}\Bigr )=0.
$$
Calculating the residues and multiplying by $\mu$, we obtain:
\beq\label{A6}
\begin{array}{c}
\nu (\mu - \nu )\tau
\Bigl ({\bf t}-[\nu ^{-1}]_{\alpha}\Bigr )
\tau \Bigl ({\bf t}+ [\mu ^{-1}]\Bigr )-\nu (\mu - \nu )
\tau ({\bf t})\tau \Bigl ({\bf t}+ [\mu ^{-1}]-[\nu ^{-1}]_{\alpha}\Bigr )
\\ \\
-\nu \p_{t_{\alpha , 1}}\tau ({\bf t}) 
\tau \Bigl ({\bf t}+ [\mu ^{-1}]-[\nu ^{-1}]_{\alpha}\Bigr )+\nu
\tau ({\bf t}) \p_{t_{\alpha , 1}}
\tau \Bigl ({\bf t}+ [\mu ^{-1}]-[\nu ^{-1}]_{\alpha}\Bigr )
\\ \\
\displaystyle{-\sum_{\gamma \neq \alpha}
\tau_{\alpha \gamma} ({\bf t})
\tau_{\gamma \alpha} \Bigl ({\bf t}+ [\mu ^{-1}]-[\nu ^{-1}]_{\alpha}\Bigr )=0.}
\end{array}
\eeq
Note that under the identification $\tau_{\alpha \beta}^p=\tau_{\alpha \beta}
\Bigl ({\bf t}-p[\mu ^{-1}]\Bigr )$ and the formal 
substitution $\mu =0$ equations (\ref{A5}) and (\ref{A6}) become the corresponding 
bilinear equations for the matrix modified KP hierarchy \cite{Z18}. 

\subsection*{Derivation of the linear problems (\ref{l1a}), (\ref{l2a})}

Here we show that the linear problems (\ref{l1a}), (\ref{l2a}), which are the basic tools
for deriving the equations of motion for the discrete time 
pole dynamics, are equivalent to corollaries
of the bilinear identity (\ref{m5}). The derivation is similar to the one given in \cite{Z18}.

Let us give some details of the calculations for the linear problem (\ref{l1a}). 
We start from the case $\alpha \neq \beta$. Substituting
$$
\Psi_{\alpha \beta}^p=\epsilon_{\alpha \beta}\frac{\tau_{\alpha \beta}^p
({\bf t}-[z^{-1}]_{\beta})}{\tau^p ({\bf t})}\, z^{\delta_{\alpha \beta}-1}
\Bigl (1-\frac{z}{\mu}\Bigr )^p e^{xz+\xi ({\bf t}, z)}
$$
and
$$
w_{\alpha \gamma}^{(1)}(p)=
\left \{
\begin{array}{l}
\displaystyle{\epsilon_{\alpha \gamma}\, \frac{\tau_{\alpha \gamma}^p
({\bf t})}{\tau^p ({\bf t})}}\qquad
\,\,\,\,\, \mbox{if $\alpha \neq \gamma$}
\\ \\
\displaystyle{-\, \frac{\p_{t_{\alpha , 1}}\tau^p ({\bf t})}{\tau^p ({\bf t})}} \qquad 
\mbox{if $\alpha = \gamma$}
\end{array}\right.
$$
into (\ref{l1a}), we write it in the form
$$
\mu z^{-1}\epsilon_{\alpha \beta} \frac{\tau_{\alpha \beta}^p
({\bf t}-[z^{-1}]_{\beta})}{\tau^p({\bf t})}+
(1-\mu z^{-1})\, \epsilon_{\alpha \beta}
\frac{\tau_{\alpha \beta}^{p+1}
({\bf t}-[z^{-1}]_{\beta})}{\tau^{p+1}({\bf t})}
$$
$$
=\, \epsilon_{\alpha \beta} \frac{\tau_{\alpha \beta}^p
({\bf t}-[z^{-1}]_{\beta})}{\tau^p({\bf t})}+z^{-1}\epsilon_{\alpha \beta}
\, \p_{t_1}\! \left (\frac{\tau_{\alpha \beta}^p
({\bf t}-[z^{-1}]_{\beta})}{\tau^p({\bf t})}\right )
$$
$$
+\sum_{\gamma \neq \alpha} z^{\delta_{\gamma \beta}-1}
\epsilon_{\alpha \gamma}\epsilon_{\gamma \beta}\left (
\frac{\tau_{\alpha \gamma}^{p+1}
({\bf t}}{\tau^{p+1}({\bf t})}-\frac{\tau_{\alpha \gamma}^{p}
({\bf t}}{\tau^{p}({\bf t})}\right )
\frac{\tau_{\gamma \beta}^p
({\bf t}-[z^{-1}]_{\beta})}{\tau^p({\bf t})}
$$
$$
-z^{-1}\epsilon_{\alpha \beta}\left (\frac{\p_{t_{\alpha , 1}}
\tau^{p+1}({\bf t})}{\tau^{p+1}({\bf t})}-
\frac{\p_{t_{\alpha , 1}}
\tau^{p}({\bf t})}{\tau^{p}({\bf t})}\right )
\frac{\tau_{\alpha \beta}^p
({\bf t}-[z^{-1}]_{\beta})}{\tau^p({\bf t})}.
$$
After some obvious transformations, separating the terms with the denominator
$(\tau^p({\bf t}))^2$, we can rewrite this as
$$
-(1-\mu z^{-1})\epsilon_{\alpha \beta}
\frac{\tau_{\alpha \beta}^{p+1}({\bf t}-[z^{-1}]_{\beta})}{\tau^{p+1}({\bf t})}+
(1-\mu z^{-1})\epsilon_{\alpha \beta}
\frac{\tau_{\alpha \beta}^{p}({\bf t}-[z^{-1}]_{\beta})}{\tau^{p}({\bf t})}
$$
$$
+\epsilon_{\alpha \beta}
\frac{\tau_{\alpha \beta}^{p+1}({\bf t})
\tau^p({\bf t}-[z^{-1}]_{\beta})}{\tau^{p+1}({\bf t})\tau^p({\bf t})}
+z^{-1}\epsilon_{\alpha \beta}
\frac{\p_{t_1}\tau_{\alpha \beta}^p({\bf t}-[z^{-1}]_{\beta})}{\tau^p({\bf t})} -
z^{-1}\epsilon_{\alpha \beta}\frac{\p_{t_{\alpha , 1}}\tau^{p+1}({\bf t})
\tau_{\alpha \beta}^p({\bf t}-[z^{-1}]_{\beta})}{\tau^{p+1}({\bf t})\tau^p({\bf t})}
$$
$$
+z^{-1}\! \sum_{\gamma \neq \alpha , \beta}
\epsilon_{\alpha \gamma}\epsilon_{\gamma \beta}\,
\frac{\tau_{\alpha \gamma}^{p+1}({\bf t})
\tau_{\gamma \beta}^{p}({\bf t}-[z^{-1}]_{\beta})}{\tau^{p+1}({\bf t})\tau^p({\bf t})}
$$
$$
+\frac{\epsilon_{\alpha \beta}}{(\tau^p ({\bf t}))^2}\left \{
-z^{-1}\tau_{\alpha \beta}^p({\bf t}-[z^{-1}]_{\beta})\p_{t_1}\tau^p({\bf t})
-\tau_{\alpha \beta}^p ({\bf t})\tau^p({\bf t}-[z^{-1}]_{\beta})
\phantom{\sum_{\gamma \neq \alpha }^N}
\right.
$$
$$
\left.  + z^{-1}\p_{t_{\alpha , 1}}\tau^p({\bf t})\,
\tau_{\alpha \beta}^{p}({\bf t}-[z^{-1}]_{\beta})-
z^{-1}\!\!\sum_{\gamma \neq \alpha , \beta}
\frac{\epsilon_{\alpha \gamma}\epsilon_{\gamma \beta}}{\epsilon_{\alpha \beta}}\,
\tau_{\alpha \gamma}^p({\bf t})\tau_{\gamma \beta}^p({\bf t}-[z^{-1}]_{\beta})\right \}=0.
$$
Th idea is to transform the expression in the brackets $\{\ldots \}$ using the 
bilinear relations (\ref{A1}) and (\ref{A2}). Namely, taking into account that
$\p_{t_1}=\sum_{\gamma}\p_{t_{\gamma , 1}}$, we rewrite it in the form
$$
\left \{\phantom{\int}\!\!\!\! \ldots \phantom{\int}\!\!\!\! \right \}=
-z^{-1} 
\tau_{\alpha \beta}^p \Bigl ({\bf t}-[z^{-1}]_{\beta}\Bigr )
\p_{t_{\beta , 1}}\tau^p({\bf t})-
\tau_{\alpha \beta}^p \Bigl ({\bf t})\tau^p({\bf t}-[z^{-1}]_{\beta}\Bigr )
$$
$$
-z^{-1}\!\! \sum_{\gamma \neq \alpha , \beta}\left [
\tau_{\alpha \beta}^p \Bigl ({\bf t}-[z^{-1}]_{\beta}\Bigr )
\p_{t_{\gamma , 1}}\tau^p({\bf t})+
\frac{\epsilon_{\alpha \gamma}\epsilon_{\gamma \beta}}{\epsilon_{\alpha \beta}}\,
\tau_{\alpha \gamma}^p({\bf t})\tau_{\gamma \beta}^p \Bigl ({\bf t}-[z^{-1}]_{\beta}\Bigr )
\right ]
$$
and apply (\ref{A2}) with $\mu =\infty$, $\nu =z$ in the first line and 
(\ref{A1}) with $\mu =z$ in the second line. The result is
$$
\left \{\phantom{\int}\!\!\!\! \ldots \phantom{\int}\!\!\!\! \right \}=
-\tau^p({\bf t})\left [z^{-1}\! \sum_{\gamma \neq \alpha}\p_{t_{\gamma , 1}}
\tau_{\alpha \beta}^p \Bigl ({\bf t}-[z^{-1}]_{\beta}\Bigr )+
\tau_{\alpha \beta}^p \Bigl ({\bf t}-[z^{-1}]_{\beta}\Bigr )\right ].
$$
Substituting this back and multiplying by $\tau^{p+1}({\bf t})\tau^p({\bf t})$,
we obtain, after some cancellations:
$$
z\epsilon_{\alpha \beta}\tau_{\alpha \beta}^{p+1}({\bf t})
\tau^p \Bigl ({\bf t}-[z^{-1}]_{\beta}\Bigr )-
z\epsilon_{\alpha \beta}\tau^{p}({\bf t})
\tau_{\alpha \beta}^{p+1} \Bigl ({\bf t}-[z^{-1}]_{\beta}\Bigr )
$$
$$
+\mu \epsilon_{\alpha \beta}\tau^{p}({\bf t})
\tau_{\alpha \beta}^{p+1} \Bigl ({\bf t}-[z^{-1}]_{\beta}\Bigr )-
\mu \epsilon_{\alpha \beta} \tau^{p+1}({\bf t})
\tau_{\alpha \beta}^{p} \Bigl ({\bf t}-[z^{-1}]_{\beta}\Bigr )
$$
$$
+\epsilon_{\alpha \beta}\p_{t_{\alpha , 1}}
\tau_{\alpha \beta}^{p} \Bigl ({\bf t}-[z^{-1}]_{\beta}\Bigr )
\tau^{p+1}({\bf t})
-\epsilon_{\alpha \beta}\p_{t_{\alpha , 1}}\tau^{p+1}({\bf t})
\tau_{\alpha \beta}^{p} \Bigl ({\bf t}-[z^{-1}]_{\beta}\Bigr )
$$
$$
+\sum_{\gamma \neq \alpha , \beta} \epsilon_{\alpha \gamma}
\epsilon_{\gamma \beta}\tau_{\alpha \gamma}^{p+1}({\bf t})
\tau_{\gamma \beta}^{p} \Bigl ({\bf t}-[z^{-1}]_{\beta}\Bigr )=0.
$$
Taking into account that $\tau_{\alpha \beta}^p ({\bf t})=
\tau_{\alpha \beta}\Bigl ({\bf t}-p[\mu^{-1}]\Bigr )$, one can see that this
is exactly the bilinear relation (\ref{A5}), where one should put $\nu =z$.

Let us now pass to the case $\alpha = \beta$ in (\ref{l1a}):
$$
\mu \, \frac{\tau^{p}({\bf t}-[z^{-1}]_{\alpha})}{\tau^{p}({\bf t})}+
(z-\mu ) \frac{\tau^{p+1}({\bf t}-[z^{-1}]_{\alpha})}{\tau^{p+1}({\bf t})}
=z\, \frac{\tau^{p}({\bf t}-[z^{-1}]_{\alpha})}{\tau^{p}({\bf t})}+
\p_{t_1}\! \left 
(\frac{\tau^{p}({\bf t}-[z^{-1}]_{\alpha}}{\tau^{p}({\bf t})}\right )
$$
$$
-z^{-1}\sum_{\gamma \neq \alpha}\left (
\frac{\tau_{\alpha \gamma}^{p+1}({\bf t})}{\tau^{p+1}({\bf t})}-
\frac{\tau_{\alpha \gamma}^{p}({\bf t})}{\tau^{p}({\bf t})}\right )
\frac{\tau_{\gamma \alpha}^{p}({\bf t}-[z^{-1}]_{\alpha})}{\tau^{p}({\bf t})}
$$
$$
-\left (
\frac{\p_{t_{\alpha , 1}}\tau^{p+1}({\bf t})}{\tau^{p+1}({\bf t})}-
\frac{\p_{t_{\alpha , 1}}\tau^{p}({\bf t})}{\tau^{p}({\bf t})}\right )
\frac{\tau^p ({\bf t}-[z^{-1}]_{\alpha})}{\tau^p({\bf t})}.
$$
Separating the terms with the denominator
$(\tau^p({\bf t}))^2$, we rewrite this as
$$
(\mu -z)\, \frac{\tau^{p+1}({\bf t}-[z^{-1}]_{\alpha})}{\tau^{p+1}({\bf t})}
-(\mu -z)\, \frac{\tau^{p}({\bf t}-[z^{-1}]_{\alpha})}{\tau^{p}({\bf t})}
+\frac{\p_{t_1}\tau^{p}({\bf t}-[z^{-1}]_{\alpha})}{\tau^{p}({\bf t})}
$$
$$
-z^{-1}\sum_{\gamma \neq \alpha}
\frac{\tau_{\alpha \gamma}^{p+1}({\bf t})
\tau_{\gamma \alpha}^{p}({\bf t}-[z^{-1}]_{\alpha})}{\tau^{p+1}({\bf t})\tau^p({\bf t})}
-\frac{\p_{t_{\alpha , 1}}\tau^{p+1}({\bf t})
\tau^p ({\bf t}-[z^{-1}]_{\alpha})}{\tau^{p+1}({\bf t})\tau^p({\bf t})}
$$
$$
+\frac{1}{(\tau^p({\bf t}))^2}
\left \{\sum_{\gamma \neq \alpha}\Bigl (
z^{-1}\tau_{\alpha \gamma}^p({\bf t})\tau_{\gamma \alpha}^p
\Bigl ({\bf t}-[z^{-1}]_{\alpha}\Bigr )
-\tau^p \Bigl ({\bf t}-[z^{-1}]_{\alpha}\Bigr )\p_{t_{\gamma , 1}}\tau^p({\bf t})
\Bigr )
\right \}=0.
$$
Using the 3-term relation (\ref{A4}), we have for the expression in the brackets 
$\{\ldots \}$:
$$
\left \{\phantom{\int}\!\!\!\! \ldots \phantom{\int}\!\!\!\! \right \}=
-\tau^p({\bf t})\sum_{\gamma \neq \alpha}\p_{t_{\gamma , 1}}
\tau^p \Bigl ({\bf t}-[z^{-1}]_{\alpha}\Bigr ),
$$
so, after multiplying by $\tau^{p+1}({\bf t})\tau^p({\bf t})$, we obtain for the 
previous expression:
$$
z(\mu -z)\tau^p({\bf t})\tau^{p+1}\Bigl ({\bf t}-[z^{-1}]_{\alpha}\Bigr )-
z(\mu -z)\tau^{p+1}({\bf t})\tau^{p}\Bigl ({\bf t}-[z^{-1}]_{\alpha}\Bigr )
$$
$$
+z\p_{t_{\alpha , 1}}\tau^{p}\Bigl ({\bf t}-[z^{-1}]_{\alpha}\Bigr )
\tau^{p+1}({\bf t})-z\p_{t_{\alpha , 1}}\tau^{p+1}({\bf t})
\tau^{p}\Bigl ({\bf t}-[z^{-1}]_{\alpha}\Bigr )
$$
$$
-\sum_{\gamma \neq \alpha}\tau_{\alpha \gamma}^{p+1}({\bf t})
\tau_{\gamma \alpha}^{p}\Bigl ({\bf t}-[z^{-1}]_{\alpha}\Bigr )=0.
$$
One can see that this is exactly the bilinear relation (\ref{A6}), where
one should put $\nu =z$. 

Equation (\ref{l2a}) for $\Psi^{\dag}$ can be processed in a similar way using
the bilinear relations (\ref{A1}), (\ref{A3}), (\ref{A4}). 

\section*{Acknowledgments}

This work was funded by the Russian Academic Excellence
Project `5-100' and supported in part by RFBR grant
18-01-00461.

\end{document}